\documentstyle[preprint,aps,prl,epsf]{revtex}

\newcommand{\sigmattbar}{\sigma _{t\overline{t}}\ }
\newcommand{\pt}{p_T}
\newcommand {\ebar}{\hbox{E\kern-0.5em\lower-0.1ex\hbox{/}}}
\newcommand {\BRtHb}{{\cal B}^{\rm t}_{\rm Hb} }
\newcommand {\BRHtaunu}{{\cal B}^{\rm H}_{\rm \tau\nu} }
\newcommand{\met}{\mbox{${\rm \not\! E}_{t}$}}
\newcommand {\WWbbbar}{{W^+ W^-} b\bar{b} }
\newcommand {\WHbbbar}{{W^{\pm} H^{\mp}} b\bar{b} }
\newcommand {\HHbbbar}{{H^+ H^-} b\bar{b} }
\newcommand {\goes}{\rightarrow}
\newcommand {\Wtaunu}{W \goes \tau \nu}
\def\bbbar{b{\bar b}}
\def\ttbar{t{\bar t}}
\def\D0{D\O} 
\begin{document}
\draft
\title{
\vspace*{-1.8cm}
\begin{flushright} 
\end{flushright}
Search for the Charged Higgs Boson in the Decays of Top Quark Pairs
in the $e\tau$ and $\mu\tau$ Channels at $\sqrt{s}$=1.8 TeV}
\font\eightit=cmti8
\def\r#1{\ignorespaces $^{#1}$}
\author{ 
\hfilneg
\begin{sloppypar}
\noindent   
T.~Affolder,\r {21} H.~Akimoto,\r {43}
A.~Akopian,\r {36} M.~G.~Albrow,\r {10} P.~Amaral,\r 7 S.~R.~Amendolia,\r {32} 
D.~Amidei,\r {24} K.~Anikeev,\r {22} J.~Antos,\r 1 
G.~Apollinari,\r {36}
% ...
\end{sloppypar}
\vskip .026in
\begin{center}
(CDF Collaboration)
\end{center}
\vskip .026in
\begin{center}
\r 1  {\eightit Institute of Physics, Academia Sinica, Taipei, Taiwan 11529, 
Republic of China} \\
\r 2  {\eightit Argonne National Laboratory, Argonne, Illinois 60439} \\
\r 3  {\eightit Istituto Nazionale di Fisica Nucleare, University of Bologna,
I-40127 Bologna, Italy} \\
\r 4  {\eightit Brandeis University, Waltham, Massachusetts 02254} \\
\r 5  {\eightit University of California at Los Angeles, Los 
Angeles, California  90024} \\  
\r 6  {\eightit Instituto de Fisica de Cantabria, University of Cantabria, 
39005 Santander, Spain} \\
\r 7  {\eightit Enrico Fermi Institute, University of Chicago, Chicago, 
Illinois 60637} \\
\r 8  {\eightit Joint Institute for Nuclear Research, RU-141980 Dubna, Russia}
\\
\r 9  {\eightit Duke University, Durham, North Carolina  27708} \\
\r {10}  {\eightit Fermi National Accelerator Laboratory, Batavia, Illinois 
60510} \\
\r {11} {\eightit University of Florida, Gainesville, Florida  32611} \\
\r {12} {\eightit Laboratori Nazionali di Frascati, Istituto Nazionale di Fisica
               Nucleare, I-00044 Frascati, Italy} \\
\r {13} {\eightit University of Geneva, CH-1211 Geneva 4, Switzerland} \\
\r {14} {\eightit Harvard University, Cambridge, Massachusetts 02138} \\
\r {15} {\eightit Hiroshima University, Higashi-Hiroshima 724, Japan} \\
\r {16} {\eightit University of Illinois, Urbana, Illinois 61801} \\
\r {17} {\eightit The Johns Hopkins University, Baltimore, Maryland 21218} \\
\r {18} {\eightit Institut f\"{u}r Experimentelle Kernphysik, 
Universit\"{a}t Karlsruhe, 76128 Karlsruhe, Germany} \\
\r {19} {\eightit Korean Hadron Collider Laboratory: Kyungpook National
University, Taegu 702-701; Seoul National University, Seoul 151-742; and
SungKyunKwan University, Suwon 440-746; Korea} \\
\r {20} {\eightit High Energy Accelerator Research Organization (KEK), Tsukuba, 
Ibaraki 305, Japan} \\
\r {21} {\eightit Ernest Orlando Lawrence Berkeley National Laboratory, 
Berkeley, California 94720} \\
\r {22} {\eightit Massachusetts Institute of Technology, Cambridge,
Massachusetts  02139} \\   
\r {23} {\eightit Institute of Particle Physics: McGill University, Montreal 
H3A 2T8; and University of Toronto, Toronto M5S 1A7; Canada} \\
\r {24} {\eightit University of Michigan, Ann Arbor, Michigan 48109} \\
\r {25} {\eightit Michigan State University, East Lansing, Michigan  48824} \\
\r {26} {\eightit University of New Mexico, Albuquerque, New Mexico 87131} \\
\r {27} {\eightit The Ohio State University, Columbus, Ohio  43210} \\
\r {28} {\eightit Osaka City University, Osaka 588, Japan} \\
\r {29} {\eightit University of Oxford, Oxford OX1 3RH, United Kingdom} \\
\r {30} {\eightit Universita di Padova, Istituto Nazionale di Fisica 
          Nucleare, Sezione di Padova, I-35131 Padova, Italy} \\
\r {31} {\eightit University of Pennsylvania, Philadelphia, 
        Pennsylvania 19104} \\   
\r {32} {\eightit Istituto Nazionale di Fisica Nucleare, University and Scuola
               Normale Superiore of Pisa, I-56100 Pisa, Italy} \\
\r {33} {\eightit University of Pittsburgh, Pittsburgh, Pennsylvania 15260} \\
\r {34} {\eightit Purdue University, West Lafayette, Indiana 47907} \\
\r {35} {\eightit University of Rochester, Rochester, New York 14627} \\
\r {36} {\eightit Rockefeller University, New York, New York 10021} \\
\r {37} {\eightit Rutgers University, Piscataway, New Jersey 08855} \\
\r {38} {\eightit Texas A\&M University, College Station, Texas 77843} \\
\r {39} {\eightit Texas Tech University, Lubbock, Texas 79409} \\
\r {40} {\eightit Istituto Nazionale di Fisica Nucleare, University of Trieste/
Udine, Italy} \\
\r {41} {\eightit University of Tsukuba, Tsukuba, Ibaraki 305, Japan} \\
\r {42} {\eightit Tufts University, Medford, Massachusetts 02155} \\
\r {43} {\eightit Waseda University, Tokyo 169, Japan} \\
\r {44} {\eightit University of Wisconsin, Madison, Wisconsin 53706} \\
\r {45} {\eightit Yale University, New Haven, Connecticut 06520} \\
\end{center}
}

\maketitle

\vspace{-0.6cm}

\begin{abstract}

Top quark production offers the unique opportunity to search  for a charged
Higgs boson ($H^\pm$), as the contribution from $t\rightarrow H^+b\rightarrow
\tau^+\nu b$ can be large in extensions of the Standard Model. We use results
from a search for top quark pair production by the Collider Detector at Fermilab
(CDF) in the e$\tau + \ebar_T$+jets and $\mu\tau +\ebar_T$+jets signatures 
to set an upper limit on the branching ratio of  
${\cal B} (t \rightarrow H^{+} b)$ in 106 pb$^{-1}$ of data. 
The upper limit is
in the range  0.5 to 0.6 at 95\% C.L. for  $H^{+}$ masses in the range 60 to
160 GeV,  assuming the branching ratio  for \mbox{$H^+\rightarrow \tau \nu$} is
100\% . The $\tau$ lepton is detected through its 1-prong  and 3-prong hadronic
decays. 

\end{abstract} 
%
%
Many extensions of the Standard Model (SM) include a Higgs sector with two 
Higgs doublets, resulting in the existence of charged (H$^{\pm}$) as well as
neutral (h, H$^0$, A) Higgs bosons. The simplest extensions are the Two-Higgs
Doublet Models (2HDMs) \cite{2HDM}, in which the extension consists only of the
extra doublet. 
In a Type I 2HDM only one of the Higgs doublets couples to
fermions, while in a Type II model one Higgs doublet couples to the ``up''
fermions (e.g., u,c,t), while the other couples to the ``down'' fermions (e.g.,
d,s,b). The Minimal Supersymmetric Model (MSSM)\cite{MSSM} is a further
extension of the SM, and has a Higgs sector like that of a Type II \mbox{2HDM}.

If the charged Higgs boson is lighter than the top
quark~\cite{top_PRD,top_prl,top_D0},  i.e. \mbox{$m_{H^\pm} <(m_{top} - m_b$)},
the decay mode $t\rightarrow H^+ b$ will compete with the SM decay 
$t\rightarrow W^+ b$. The consequence is that $\ttbar$ production and decay 
will provide a source of Higgs bosons in the channels  $\WHbbbar$ and $\HHbbbar$
produced with a strong--interaction  rather than the weak--interaction
cross--section of direct $H^+ H^-$ pair production.  In addition, the signature
from top pair production and decay is much cleaner than that of the direct
production with respect to QCD background.

\begin{figure}
\vspace{0cm}
\epsfxsize=7.5cm
\centerline{\epsffile{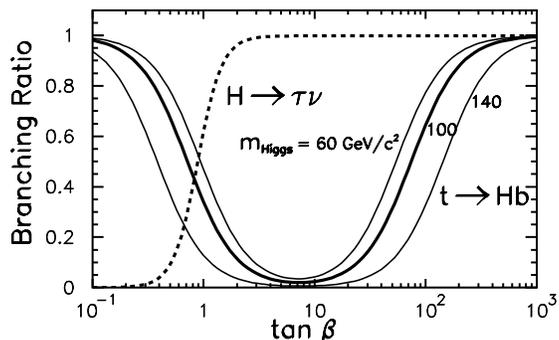}}
\caption{Branching fraction of $H\rightarrow\tau\nu$ and $t\rightarrow Hb$ as a
function of $\tan\beta$ at lowest order in the MSSM. The top quark mass is
assumed to be 175 GeV/c$^2$.}
\label{fig:br}
\end{figure}

In a 2HDM and in the MSSM the branching ratio for \mbox{$t\rightarrow H^+ b$},
$\BRtHb$,
depends on the charged Higgs mass and $\tan\beta$, the ratio of the vacuum
expectation values for the two Higgs doublets. Figure~\ref{fig:br} shows the
expected branching ratio from a leading--log QCD calculation~\cite{dproy_calc}
in the MSSM for three different charged Higgs masses $m_{H^\pm} = 60, 100, 140$
GeV/$c^2$ as a function of $\tan\beta$. For  \mbox{ $\tan\beta$
\raisebox{-.5ex}{$\stackrel{\textstyle <}{\sim}$} 1} and  \mbox{
$\tan\beta$\raisebox{-.5ex}{$\stackrel{\textstyle >}{\sim}$} 70} the MSSM 
predicts that the decay mode $t\rightarrow H^+ b$ dominates. 
Also shown in Figure~\ref{fig:br} is the predicted branching ratio in the MSSM
at lowest order for the decay of the charged Higgs boson into a charged
$\tau$--lepton and a $\tau$--neutrino ($\BRHtaunu$), which has little dependence
on the charged Higgs mass. For $\tan\beta > 1$ the decay
$H^+\rightarrow\tau^+\nu_\tau$ is predicted to dominate over the other main
decay mode, $H^+\rightarrow c\overline{s}$, and for $\tan\beta > 5$ the 
branching ratio $\BRHtaunu$ is expected to be nearly 100\%. Thus, this model
would  predict an excess of top events with tau leptons over the number expected
from SM events in which $\ttbar \goes \WWbbbar$, followed by $\Wtaunu$.

Recent calculations, however, have shown that at large values of $\tan\beta$
the predicted  branching ratio for
$t \rightarrow H^+b$ is highly sensitive
to higher--order radiative  corrections,
which are model--dependent~\cite{Coarasa_Guasch_Sola}.  Limits in the
$\tan\beta - m_{H^\pm}$ parameter plane consequently depend critically on the
parameters of the model. However the direct search for the signature of a $\tau$
lepton in top decays allows us to set an upper limit on the 
branching ratio of \mbox{$t \rightarrow H^+ b$}, assuming the branching ratio 
for \mbox{$H^+\rightarrow \tau \nu$} is 100\%, for example.

Previous searches for the charged Higgs boson in top decay have been in the
$\tau+\met$ channel~\cite{eta}\cite{UA1UA2}, $\ell\ell+\ebar_T$+X ($\ell =$ e or $\mu$,
X=anything) channel~\cite{JinSong}, the $\ebar_T +\tau$+jets
channel~\cite{CJessop,CCouyumt},  the $\ell$ +jets channel~\cite{D0,Bevensee},
and the  $\ebar_T+\tau b + O$+ jet ($O=e$, $\mu$, $\tau$ or jet)
channel~\cite{Conway}. Both Ref.~\cite{CJessop} and Ref.~\cite{Conway} select
events with  a $\ebar_T$ trigger, while Ref.~\cite{D0,Bevensee} is an
indirect search using a  disappearance method. Searches for direct production at
LEP set a lower limit on the mass of 69 GeV/c$^2$\cite{LEP}. Indirect limits
have also been set from  measurements of the rate for the decay $b \rightarrow
s\gamma$\cite{Bsgamma}.  However higher--order calculations have shown that in
both 2HDM models\cite{Borzumati_2HDM} and the  MSSM\cite{CGHSola}  these limits
are also highly model--dependent. 

The CDF collaboration has published a search for $\tau$ leptons from decays of top quark pairs in
the $\ell\tau+\ebar_T+$2 jets$+$X $(\ell=e,\mu)$ channel~\cite{our_prl}, where
events were selected by requiring the presence of a high--p$_T$ $e$ or $\mu$. We present
here the constraints that this analysis (the``$\ell\tau$" analysis)  imposes on
the branching ratio of the top quark into a  charged Higgs boson. This was 
suggested in Ref.~\cite{dproy_prd}, where the authors compare the CDF data with
a generator--level Monte Carlo calculation for the number of expected events
from charged Higgs decay. 

In this paper we start with the number of top candidate events found in the 
$\ell\tau+\ebar_T+2$ jets$+$X data in the analysis of Ref.~\cite{our_prl}.
We then apply the same selection criteria 
to Monte Carlo events that contain top quark 
pairs in which one or both top quarks 
decay to the charged Higgs (i.e. $\ttbar\rightarrow \WHbbbar$ 
and $\ttbar\rightarrow \HHbbbar$), for different Higgs masses. 
We assume there are no top quark decays other than $t\rightarrow
W^+ b$ and  $t\rightarrow H^+ b$.
We perform a full calculation of the acceptances including
detector effects, and determine the expected number of events due to 
Higgs production and subsequent decay. From this we can set a limit on 
the branching ratio $t \rightarrow H^+b$. 

The selection used in this analysis requires \mbox{high--$p_T$} inclusive lepton
events that contain an electron with $E_T>20$ GeV or a muon with $p_T>20$
GeV/$c$ in the central region ($|\eta|<1.0$). The other lepton  must be a tau
lepton, also in the central region, with  momentum $p_T>15$
GeV/$c$\cite{pt_fudge}. Dilepton events from $\ttbar$ decays are expected to
contain two jets from $b$ decays and large missing transverse energy from the
neutrinos. Therefore, we select events with $\geq 2$ jets (with $E_T > 10$ GeV
and \mbox{$|\eta| < 2.0$}), and with large $\ebar_T$ significance
($S_{E\!\!\!\!/_T}>3$),  as described in detail in Ref.~\cite{our_prl}.

Two complementary techniques, one which identified the $\tau$ lepton starting
with clusters in the calorimeter, and another which started with a high $\pt$
single track, were used for identifying hadronically decaying 
$\tau$'s\cite{our_prl}.  Here, we combine the two tau selections by accepting
events which pass either set of criteria. Both techniques find the same four top
dilepton candidates in 106~pb$^{-1}$ of data. 
The total acceptance of the combined selection for SM top quark pairs decays, i.e. the
events  that pass the final cuts divided by the number of generated $\ttbar$ events,
is (0.172$\pm$0.014)\%.
We expect a total of 3.1$\pm$0.5 events from background sources.
The dominant background is due to $Z/\gamma\rightarrow\tau^+\tau^-$+jets events 
($1.8\pm 0.5$ events), and to $W+\geq 3$~jets events where one jet is misidentified 
as a $\tau$ lepton ($1.0\pm 0.1$ events). We expect $0.3\pm 0.1$ background 
events from $WW$ and $WZ$ production.
We calculate the number of expected events in the $\ell\tau$ channel by combining
the $\ttbar$ cross section, the luminosity and the total acceptance.
For the $\ttbar$ cross section we use the CDF measurement in the ``lepton+jets''
channel, where one $W$ decays leptonically and the other $W$ decays hadronically.
This yields the most precise determination of the $\ttbar$ cross section in a single channel,
$\sigma_{t\overline{t}}=5.1\pm$1.5~pb~\cite{Paolo}.
Using this cross section we 
expect 0.9~$\pm$~0.1 events from SM $\ttbar$ decay in the $e\tau$ and $\mu\tau$ channels.

Although  the identification of $b$ quarks was not part
of the search criteria, three of the four candidate events contain at least one
$b$--tagged jet\cite{btag}, 
while we expect 0.2 tagged events from SM non--$\ttbar$
background~\cite{our_prl}. In the following we
will use the combined tau selection for our results. 

If a charged Higgs boson is present all three of the final states $\WWbbbar$, 
$\WHbbbar$, and $\HHbbbar$ can contribute to the $\ell\tau$ channel.  The
total acceptance for top decay in the $\ell\tau$ channel is given by

\begin{eqnarray}
A_{tot}^{\ell\tau}&=& (1-\BRtHb ) ^2 A_{WW}^{\ell\tau}+  \nonumber \\
& &2(1-\BRtHb)\BRtHb\BRHtaunu A_{WH}^{\ell\tau}+\nonumber\\
& &  (\BRtHb) ^2 (\BRHtaunu) ^2 A_{HH}^{\ell\tau} ~~.
\end{eqnarray}

Here $A_{WW}^{\ell\tau}$ is the total acceptance of the event selection
criteria for the case where the $\ttbar$ pair decays into $\WWbbbar$. It
includes the geometric and kinematic acceptances, the efficiencies for the
trigger,  lepton identification, and cuts on the event topology,  and all
branching ratios of both the $\tau$ and the $W$ boson\cite{explain_br}.
Similarly,  $A_{WH}^{\ell\tau}$ and $A_{HH}^{\ell\tau}$ are the respective
total acceptances for the $\ttbar$ pair decays into  $\WHbbbar$ and $\HHbbbar$,
but where the branching ratio of the top to Higgs ($\BRtHb$) and of the Higgs to
tau ($\BRHtaunu$) have been factored out explicitly.  We assume that $\BRHtaunu$
is 100\%, as it would be at large $\tan\beta$ in the MSSM, and set a limit on
$\BRtHb$.

We use a top quark mass of 175~GeV$/c^2$.
Monte Carlo simulations of $t\overline{t}$ production
and decay in the three modes $\WWbbbar$, $\WHbbbar$, and
$\HHbbbar$ provide estimates of the geometric and kinematic acceptance,
$A_{geom\cdot P_T}$, and of the efficiency of the cuts on the event topology
for different Higgs masses
($m_{H^\pm}=60, 80, 100, 120, 140, 160$ GeV$/c^2$). We use the {\sc
pythia}~\cite{pythia} Monte Carlo to generate $t\overline{t}$ events, the {\sc
tauola} package~\cite{tauola}, which  correctly treats the $\tau$ polarization,
to decay the tau lepton, and a detector simulation. The selection of events is
identical to that described in detail in Ref.~\cite{our_prl}. The efficiencies
for electron and muon identification are measured from 
$Z^\circ\rightarrow e^+e^-$ 
and $Z^\circ\rightarrow \mu^+\mu^-$ data.

\begin{figure}[!ht]
\vspace{0cm}
\epsfxsize=7.0cm
\centerline{\epsffile{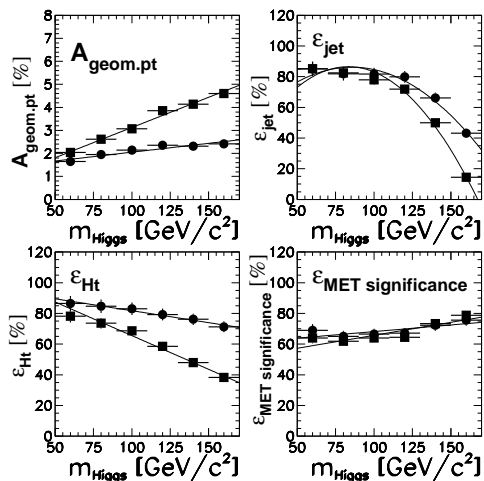}}
\caption{Contributions to the total acceptance in the $\ell\tau+\ebar+2$ jets$+X$ 
channel versus the mass of the charged Higgs boson.
The circles are for the $\WHbbbar$ decay of the $\ttbar$ pair; 
the squares for for the $\HHbbbar$ decay. 
}
\label{fig:acc_breakdown}
\end{figure}

\begin{figure}[!ht]
\vspace{0cm}
\epsfxsize=7.0cm
\centerline{\epsffile{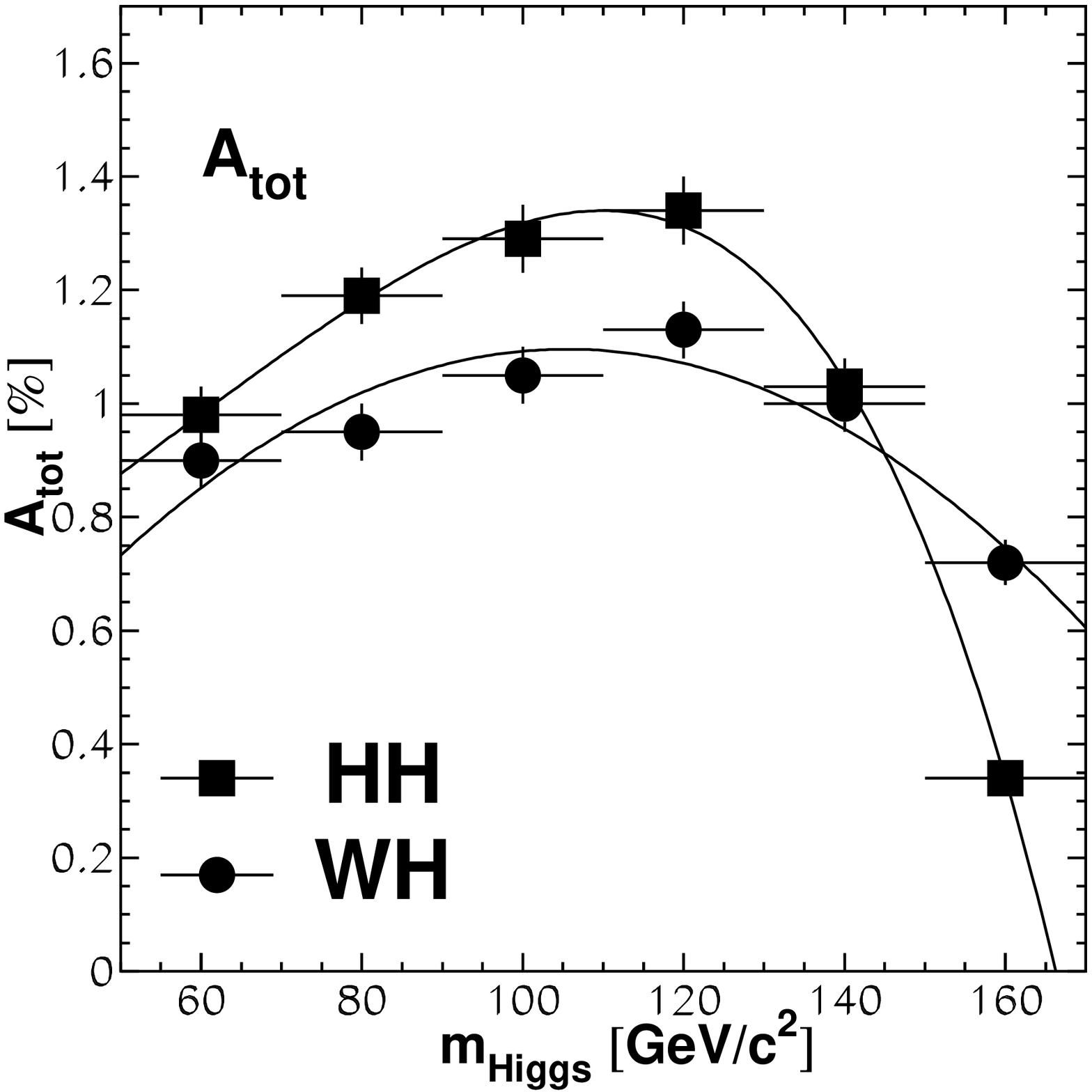}}
\caption{Total acceptance in the ``tau dilepton'' channel, versus the mass of
the Higgs boson. The circles are for the $\WHbbbar$ decay of the $\ttbar$ pair; 
the squares for the $\HHbbbar$ decay. }
\label{fig:acc_tot}
\end{figure}

Figure~\ref{fig:acc_breakdown}  shows $A_{geom\cdot P_T}$, the efficiency
$\epsilon_{jet}$ of the 2--jet cut, the efficiency $\epsilon_{H_T}$ of the cut
on the total transverse energy $H_T$~\cite{our_prl}, and the efficiency of the cut on the
$E\!\!\!\!/_T$ significance, as a function of Higgs mass. As $m_{H^\pm}$
increases the tau leptons become more energetic and  $A_{geom\cdot P_T}$
increases. When $m_{H^\pm}$ approaches $m_{top}$  the $b$ jets instead become
less energetic and $\epsilon_{jet}$ drops rapidly. Figure~\ref{fig:acc_tot}
shows the resulting values for $A_{WH}^{\ell\tau}$ and $A_{HH}^{\ell\tau}$
versus $m_{H^\pm}$; the numerical values are listed in
Table~\ref{acceptance_both}.
Note that relative to  $A_{HH}$,  $A_{WH}$ has a 
factor of 2/9 in it due to the branching ratio for $W \goes \ell \nu$, while thhe
factor of two due to the two possible charge combinations 
$W^+H^-$ and  $W^-H^+$ is explicitly included in eq. (1).
Overall, the total acceptance ($A_{tot}^{\ell\tau}$) is rather insensitive to the value of
the Higgs  mass, ranging between 0.7~\%  and 1.3~\% until $m_{H^\pm}$ approaches
$m_{top}$.  This is to be compared to the acceptance $A_{WW}^{l\tau}$ in the
$W^+W^-$ final state~\cite{our_prl} of 0.17\%.

\begin{figure}[!ht]
\vspace{0cm}
\epsfxsize=7.0cm
\centerline{\epsffile{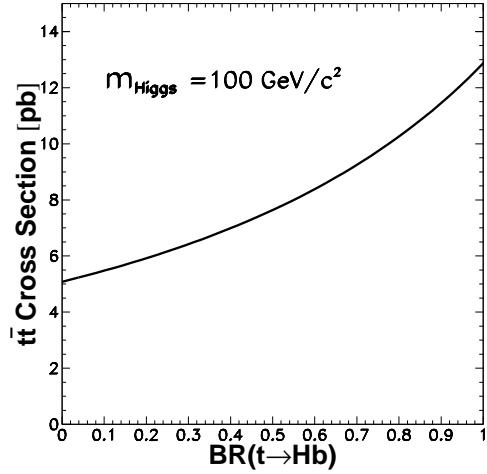}}
\caption{The $\ttbar$ cross section is a function of the branching ratio 
${\cal B} (t \rightarrow H^{+} b)$.}
\label{fig:xsec_br}
\end{figure}

\begin{figure}[!ht]
\vspace{0cm}
\epsfxsize=7.0cm
\centerline{\epsffile{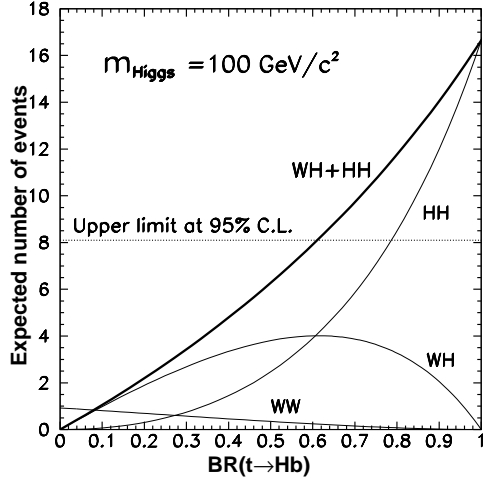}}
\caption{
The predicted number of events for 106 pb$^{-1}$ of data versus the branching
ratio for top decay into $H^{+}b$ for $m_{Higgs}$=100 GeV/c$^2$.  The graph
shows the  contributions from the $\WWbbbar$, $\WHbbbar$, and  $\HHbbbar$
channels separately.}
\label{fig:limit_br}
\end{figure}

The expected number of events in the $\ell\tau$ channel is given by
\begin{equation}
N_{exp}^{\ell\tau} = \sigma_{t\overline{t}}\cdot{\cal L}\cdot
A_{tot}^{\ell\tau}(\BRtHb ,m_{H^\pm})
\end{equation}
and depends on $\BRtHb$, the Higgs boson mass, and $\sigma_{t\overline{t}}$,
the total top pair production cross section. Rather than use the
theoretical prediction for $\sigmattbar$, 
for each value of 
$\BRtHb$ we normalize to the observed number of events in the 
`lepton + jets' channel with a secondary vertex tag, taking
into account the contributions from the three separate decay
final states of $\WWbbbar$, $\WHbbbar$, and $\HHbbbar$, calculated using
the full Monte
Carlo simulation and the updated tagging efficiency~\cite{Paolo}. 
We have checked that the calculation gives the 
value of $\sigma_{t\overline{t}}=5.1$~pb
in the SM case of $\BRtHb = 0$, in agreement with the CDF
standard model analysis of the top cross section~\cite{Paolo}, as it must be.
We thus calculate $\sigma_{t\overline{t}}$ from the number of observed events in
the lepton plus jets channel with a secondary vertex tag, $N^{\ell+jets}=29$, the
expected number of SM background events, $B^{\ell+jets}=8.0\pm 1.0$, and a total acceptance
$A_{tot}^{\ell+jets}(\BRtHb,m_{H^\pm})$ that takes the $\WHbbbar$ and
$\HHbbbar$ decay modes into account. This can be written as
\begin{eqnarray}
\sigma_{t\overline{t}} =
\frac{N^{\ell+jets} - B^{\ell+jets}}{{\cal L}\cdot A_{tot}^{\ell+jets}
(\BRtHb ,m_{H^\pm})}
\end{eqnarray}
where $A_{tot}^{\ell+jets}$ is given analogously to $A_{tot}^{\ell\tau}$
by
\begin{eqnarray}
A_{tot}^{\ell+jets} &=& (1-\BRtHb)^2 A_{WW}^{\ell+jets} + 
\nonumber \\
& &  2(1-\BRtHb)\BRtHb A_{WH}^{\ell+jets} + \nonumber\\
& &  (\BRtHb)^2 A_{HH}^{\ell+jets} ~~.
\end{eqnarray}
Figure~\ref{fig:xsec_br} shows how $\sigma_{t\overline{t}}$ increases as
$\BRtHb$ becomes larger.
The contribution from $H^+\rightarrow c\overline{s}$ decays is neglected, as we
have assumed $\BRHtaunu = $ 1. For a large branching ratio into $H^+b$,  the
$\HHbbbar$ mode becomes dominant and the leptons ($e$ or $\mu$), which in this
case originate from tau decays, have a softer $p_T$ spectrum than leptons
produced in $W$ decays, and $A_{tot}^{\ell+jets}$ decreases. 
Figure~\ref{fig:limit_br} shows the expected number of events 
versus $\BRtHb$ from each of the $\WWbbbar$, $\WHbbbar$, and $\HHbbbar$ decay
modes for $m_{H^\pm}=100$ GeV$/c^2$.

\begin{figure}[!ht]
\vspace{0cm}
\epsfxsize=7.0cm
\centerline{\epsffile{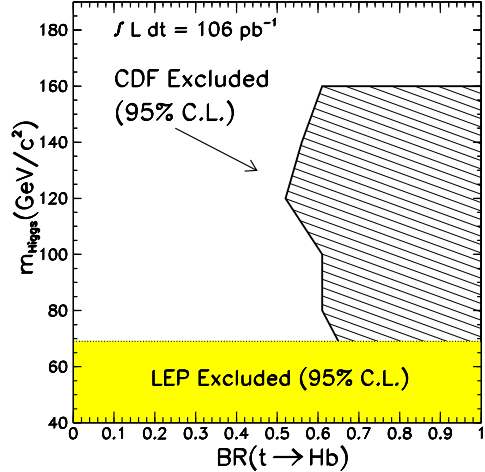}}
\caption{The region excluded at 95\% C.L. for charged Higgs production versus
the branching ratio for top decay into $H^{+}b$.}
\label{fig:exclusion_br_final}
\end{figure}

Based on the observation of 4 events and the predicted background of $3.1\pm0.5$
events, we calculate a 95\% C.L. upper limit on  Higgs production of 8.1 events.
When calculating the limit, we include the systematic uncertainties, which are
dominated by uncertainties on $N^{\ell+jets}$ (26\%), tau identification (11\%),
$b$ tagging efficiency (10\%) and Monte Carlo  statistics (8\%). Then, to
determine a limit on the branching ratio $\BRtHb$, we calculate the number of
events expected versus $\BRtHb$ for different Higgs masses in steps of 20
GeV/c$^2$. Figure~\ref{fig:exclusion_br_final} shows the 
region excluded at 95\% C.L. as a function of the branching ratio of
\mbox{$t\rightarrow H^+ b$}. The upper limit is in the range  0.5 to 0.6 at
95\% C.L. for $H^{+}$ masses in the range 60 to 160 GeV.

\begin{figure}[!ht]
\vspace{0cm}
\epsfxsize=7.0cm
\centerline{\epsffile{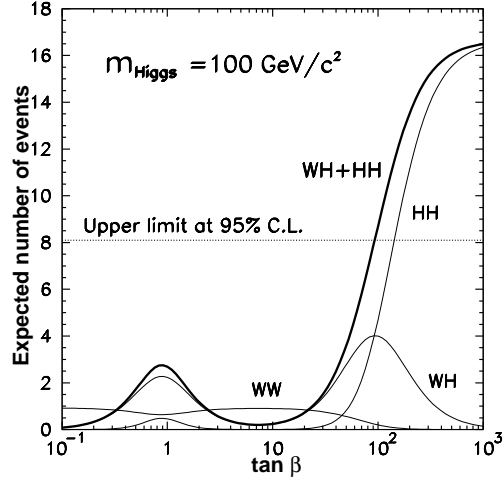}}
\caption{
The predicted number of events at lowest order in the MSSM for 106 pb$^{-1}$ of
data versus $\tan\beta$, for $m_{Higgs}$=100 GeV/c$^2$. 
The graph shows the different
contributions from the $\HHbbbar$ and $\WHbbbar$ channels separately.}
\label{fig:limit_tanbeta}
\end{figure}

\begin{figure}[!ht]
\vspace{0cm}
\epsfxsize=7.0cm
\centerline{\epsffile{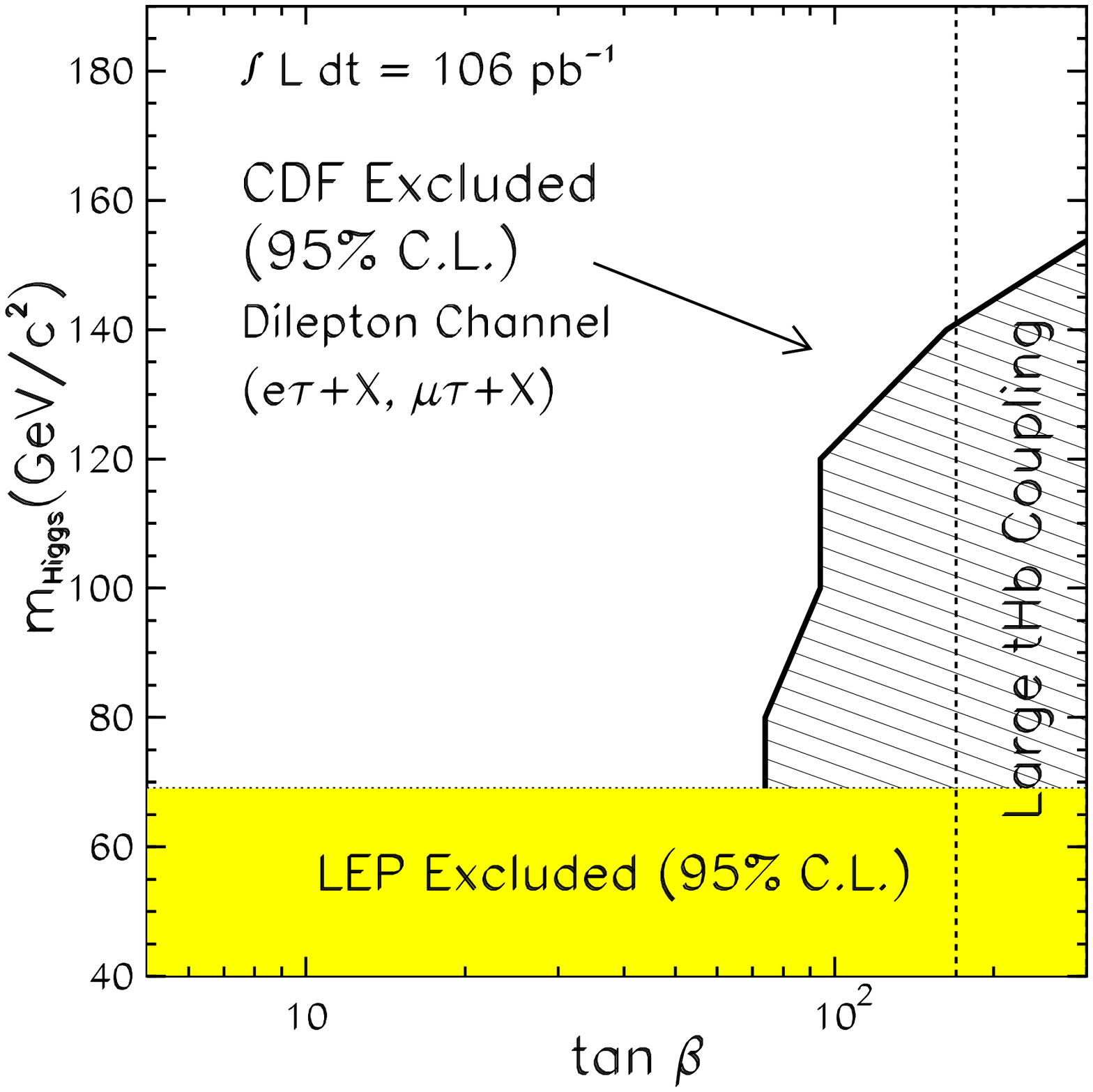}}
\caption{Excluded regions (95\% C.L.) at different values of $\tan\beta$
for charged Higgs production, at lowest order in the MSSM.
The coupling $tbH^+$ may become non--perturbative
in the region at large values of $\tan\beta$, and the limit does not apply.
}
\label{fig:exclusion_tanbeta_final}
\end{figure}

For the special case of the MSSM, although the branching ratios have been shown
to be strongly model--dependent, for the Higgs mass parameter $\mu < 0$ the SUSY
QCD  and QCD corrections come close to cancelling, and the next--to--leading 
order prediction is almost unchanged from the tree--level
result~\cite{Coarasa_Guasch_Sola}. Figure~\ref{fig:limit_tanbeta} shows the
expected number of $\ell\tau$ events  versus $\tan\beta$ from each of the
$\WWbbbar$, $\WHbbbar$, and $\HHbbbar$ decay modes for $m_{H^\pm}=100$~GeV$/c^2$, 
at lowest order in the MSSM. The shapes of the curves are mainly
due to the variation of the branching ratio $\BRtHb$ as a function of
$\tan\beta$. Figure~\ref{fig:exclusion_tanbeta_final} shows the excluded region
in the plane of $m_{H^\pm}$ and $\tan\beta$, again at lowest order in the MSSM.
In the region at large values of $\tan\beta$ the $tbH^+$ Yukawa coupling may
become non--perturbative (see  Ref.~\cite{Coarasa_Guasch_Sola}). In this case
the limit is not valid.

We compare our results to those of Ref.~\cite{dproy_prd}.  We find that the
acceptance is smaller by about a factor of two. The limits presented in this
letter use the correct $\WWbbbar$, $\WHbbbar$, and $\HHbbbar$
acceptances, including the correlations among the different objects ($e,
\mu,\tau,b$-quark) in the events.  
The insight of Ref~\cite{dproy_prd} that this will be a channel of much interest
in Fermilab Run II remains intact, however.

In conclusion, we have used the data from the CDF search\cite{our_prl}  for top
quark decays into final states containing a light lepton ($e$ or $\mu$) and a
$\tau$ lepton,  detected through its 1-prong  and 3-prong hadronic decays, to
set a limit on the branching ratio of the top quark into the charged Higgs plus
a b quark, $\BRtHb$. The limit ranges from 0.5 to 0.6 at 95\% C.L. for 
$H^{+}$ masses in the range 60 to 160 GeV,  assuming the branching ratio  for
\mbox{$H^+\rightarrow \tau \nu$} is 100\%. 

We thank D.P. Roy for stimulating our interest in
this analysis and for discussions, and G. Farrar for suggesting
that we use the \mbox{$\BRtHb - m_{H^\pm}$} plane rather than the 
\mbox{$\tan\beta - m_{H^\pm}$} plane for presenting limits.
We thank the Fermilab staff and the technical staffs of the
participating institutions for their vital contributions.  This work was
supported by the U.S. Department of Energy and National Science Foundation;
the Italian Istituto Nazionale di Fisica Nucleare; the Ministry of Education,
Science, Sports and Culture of Japan; the Natural Sciences and Engineering
Research Council of Canada; the National Science Council of the Republic of
China; the Swiss National Science Foundation; the A. P. Sloan Foundation; the
Bundesministerium fuer Bildung und Forschung, Germany; and the Korea Science
and Engineering Foundation.
%

\newpage
\begin{table*}
\begin{tabular}{ccc}
$M_{Higgs}$ & $A_{WH}^{l\tau}$ (\%) & $A_{HH}^{l\tau}$ (\%) \\ \hline
60      &   0.91$\pm$0.06    &    1.00$\pm$0.06 \\
80      &   0.98$\pm$0.06    &    1.17$\pm$0.06 \\
100     &   1.11$\pm$0.06    &    1.32$\pm$0.07 \\
120     &   1.08$\pm$0.06    &    1.32$\pm$0.07 \\
140     &   0.67$\pm$0.05    &    0.98$\pm$0.06 \\
160     &   0.72$\pm$0.05    &    0.32$\pm$0.03 \\
\end{tabular}
\caption{The total acceptance versus the mass of the charged Higgs boson for the
$\ell\tau+\ebar_T+2$ jets $+X$ analysis. The uncertainties are statistical only. These
numbers are to be compared to the acceptance for SM top quark pair decays of
$A_{WW}^{l\tau} = (0.172\pm 0.014)\% $. The larger acceptance with the charged
Higgs is primarily due to the larger branching fractions into $\tau$ leptons.}
\label{acceptance_both}
\end{table*}


\begin{thebibliography}{99}
\bibitem{eta} In the CDF coordinate system, $\theta$ and $\phi$ are the 
polar and azimuthal angles, respectively, with 
respect to the proton beam direction. The pseudorapidity $\eta$ is defined as
$-\ln\tan(\theta/2)$. The transverse momentum of a particle is $P_T = P
\sin\theta$. The analogous quantity using energies, defined as $E_T = E
\sin\theta$, is called transverse energy.
The missing transverse energy $E\!\!\!\!/_T$ is defined as $-\sum E_T^i\ 
\hat{r}_i$, where $\hat{r}_i$ are the unit vectors in the transverse plane 
pointing to the energy depositions in the calorimeter.
\bibitem{btag} We employed the same SVX + SLT b-tagging algorithms used in
Ref~\cite{top_PRD}. We emphasize again that in Ref~\cite{our_prl} (and
consequently in this paper) we did not use b--tagging in the event selection,
because of the small number of events expected from top decays.
\bibitem{explain_br} The reason the acceptance is quoted as containing the
branching ratios is that $\ttbar$ events are complex, and can satisfy the event
selection  criteria via a number of different paths, such as the $e$ or $\mu$
coming from the $W$ and the $\tau$ from the $H$ in the $WH\bbbar$ intermediate
state, or, the  $\tau$ coming from the $W$ decay and the $e$ or $\mu$ coming
from the decay of a $\tau$ from the $H$, for example. We consequently perform the
Monte Carlo calculations and include all decay modes except those involving the Higgs, for
which the  branching ratios are varied explicitly to find the acceptance for
each value. The acceptance is then defined as the number of accepted events over
the number of $\ttbar$ events generated.
\bibitem{pt_fudge} The selection criteria are that either a track exists with
$\pt > 15$ GeV/$c$ or the sum of the transverse momenta of the tracks and
the transverse energies of $\pi^0$'s be greater than  15 GeV/$c$. See 
Ref~\cite{our_prl} for details.
\end{thebibliography}
\end{document}